\documentclass[10pt]{article}
\usepackage{graphicx}
\usepackage{amsmath}
\usepackage{amssymb}
\usepackage{multirow}
\usepackage{caption2}
\usepackage[figuresright]{rotating}
\begin{document}
\bibliographystyle{prsty}
\begin{center}
{\large {\bf \sc{Strong decays of the $P_c(4312)$ and its isospin cousin via the QCD sum rules}}} \\[2mm]
Xiu-Wu Wang,
Zhi-Gang  Wang\footnote{E-mail: zgwang@aliyun.com.}\\
 Department of Physics, North China Electric Power University, Baoding 071003, P. R. China\\
\end{center}

\begin{abstract}
In the present work, considering  the conservation of isospin in the strong decays, we investigate  the strong decays of the pentaquark molecule candidate   $P_c(4312)$ and its possible higher isospin cousin $P_c(4330)$ in the framework of the QCD sum rules. What's more, the pole residue of the $\Delta$ baryon with isospin eigenstate $|II_3\rangle=|\frac{3}{2}\frac{1}{2}\rangle$ is obtained. If the possible pentaquark molecule candidate $P_c(4330)$ could be testified in the future experiment, it would shed light on interpretations  of the  $P_c$ states.
\end{abstract}



\section{Introduction}
Till now, several pentaquark candidates $P_c$ and $P_{cs}$ are observed by the LHCb collaboration:
In 2015, the LHCb collaboration observed the $P_c(4380)$ and $P_c(4450)$ via analyzing   the $\Lambda_b^0\rightarrow J/\psi K^-p$ decay \cite{RAaij1}. In 2019, the LHCb collaboration re-investigated the experimental  data with one order of magnitude larger than that previously analyzed,  and observed a narrow pentaquark candidate $P_c(4312)$ in the $J/\psi p$ mass spectrum, its measured Breit-Wigner mass and width are $4311.9\pm0.7^{+6.8}_{-0.6}\,\rm{MeV}$ and $9.8\pm2.7^{+3.7}_{-4.5}\,\rm{MeV}$, respectively \cite{RAaij2}. The investigation also showed that the $P_c(4450)$ actually consists  of two narrow overlapping peaks  $P_c(4440)$ and $P_c(4457)$ \cite{RAaij2}.

In 2020, the LHCb collaboration observed an evidence for a new  structure $P_{cs}(4459)$ in the $J/\psi \Lambda$ mass spectrum with a significance of  $3.1\sigma$ in the $\Xi_b^- \to J/\psi K^- \Lambda$ decays \cite{LHCb-Pcs4459-2012}. In 2021,  the LHCb collaboration observed an evidence for a new structure $P_c(4337)$ in the $J/\psi p$ and $J/\psi \bar{p}$ systems in the $B_s^0 \to J/\psi p \bar{p}$ decays with  a significance  of $3.1\sim3.7\, \sigma$ \cite{LHCb-Pc4337-2108}. In 2022, the LHCb collaboration observed an evidence for a new structure $P_{cs}(4338)$ in the $J/\psi \Lambda$ mass spectrum  in the $B^- \to J/\psi \Lambda \bar{p}$ decays with the  Breit-Wigner mass and width  $4338.2\pm0.7\pm0.4\,\rm{MeV}$ and $7.0\pm1.2\pm1.3\,\rm{MeV}$ respectively, and the preferred spin-parity $J^P={\frac{1}{2}}^-$ \cite{LHCb-Pcs4338}.

The spins and parities of those $P_c$ and $P_{cs}$ states have not been surely determined experimentally yet, and their physical natures are still under hot debates. Except for the $P_c(4337)$, the masses of those $P_c$ and $P_{cs}$ states are near the thresholds of the  $\Sigma_c^{(*)}\bar{D}^{(*)}$ and $\Xi_c\bar{D}^{(*)}$ pairs, respectively. Thus, a typical interpretation is that they are the S-wave hidden-charm meson-baryon molecules with definite isospin $I$, spin $J$ and parity $P$ \cite{Mass-mole-CWXiao,Mass-mole-JunHe,Mass-mole-HXChen-2,Mass-mole-HXChen,mass-mole-WZG,mass-mole-MengLin,Mass-mole-LMeng,
Mass-mole-MingZhu,Mass-JRZhang-EPJC2019,mass-mole-Azizi}. The mass and width are two basic parameters  to determine the physical nature of a hadronic state, recently, many theoretical groups applied different methods to study their masses \cite{Mass-mole-CWXiao,Mass-mole-JunHe,Mass-mole-HXChen-2,Mass-mole-HXChen,mass-mole-WZG,mass-mole-MengLin,Mass-mole-LMeng,
Mass-mole-MingZhu,Mass-JRZhang-EPJC2019,mass-mole-Azizi,mass-mole-WXW-SCPMA,mass-mole-WXW-IJMPA,mass-mole-WXW-CPC,
mass-penta-WZG-EPJC-16,mass-penta-WZG-IJMPA,mass-penta-WZG-CPC} and strong decays \cite{Decay-mole-Zou,Decay-mole-GJWang,Decay-mole-HeChen,Decay-mole-Gutsche,Decay-mole-WZG-WX,Decay-mole-HuangMQ,
Decay-mole-Sakai} under the physical picture of meson-baryon hadronic molecules.
 Different interpretations, such as the diquark-diquark-antiquark type pentaquark states and baryon-meson molecular states, lead to different branching fractions in the strong decays, moreover, even under the same physical assignments, different theoretic groups obtain very  different branching fractions. For example, in the picture of the $\bar{D}\Sigma_c$ molecular state, the calculations based on the effective Lagrangian (quasipotential Bethe-Salpeter equation) approach indicate that the $J/\psi p$ and $\eta_c p$ decay channels play a tiny role \cite{Decay-mole-Zou,Decay-mole-GJWang} (\cite{Decay-mole-HeChen}), the calculations based on the Weinberg-Salam compositeness condition indicate that the ratio $\Gamma(P_c(4312)\rightarrow \eta_cp):\Gamma(P_c(4312)\rightarrow J/\psi p) =3$ \cite{Decay-mole-Gutsche},
 the calculations based on the QCD sum rules indicate that the ratio $\Gamma(P_c(4312)\rightarrow \eta_cp):\Gamma(P_c(4312)\rightarrow J/\psi p) =0.24$ ($3.3$) \cite{Decay-mole-WZG-WX} (\cite{Decay-mole-HuangMQ}), et al.

The $P_c(4312)$ was observed in the $J/\psi p$ invariant mass spectrum, it should have the isospin $I=\frac{1}{2}$, however, the isospin of the $\bar{D}\Sigma_c$ state was not specified when it was explored with the QCD sum rules  \cite{Mass-mole-HXChen-2,Mass-mole-HXChen,mass-mole-WZG,
Mass-JRZhang-EPJC2019,mass-mole-Azizi,Decay-mole-WZG-WX,Decay-mole-HuangMQ}.
In Refs.\cite{mass-mole-WXW-SCPMA,mass-mole-WXW-IJMPA,mass-mole-WXW-CPC}, we investigate the color-singlet-color-singlet (or meson-baryon) type hidden-charm pentaquark molecular states without strange, with strange and with double-strange in a comprehensive way by distinguishing the isospin, and make possible and reasonable assignments of the $P_c(4312)$, $P_c(4380)$, $P_c(4440)$, $P_c(4457)$, $P_{cs}(4338)$, $P_{cs}(4459)$, and make many predictions on the masses for the molecular states to be confronted to the experimental data in future.

The QCD sum rules approach is a powerful non-perturbative theoretical tool proposed by Shifman, Vainshtein and Zakharov in 1979 \cite{SVZ1,SVZ2}. It has been widely applied to study the hadron's masses, decay constants, form-factors, coupling constants, etc \cite{Reinders,ColangeloReview}. We usually resort to the three-point correlation functions or two-point light-cone correlation functions to study the three-hadron coupling constants. At the QCD side, we perform the operator product expansion at the large space-like region and at the light-cone, then get the QCD spectral densities in the spectral representation and match with the hadron side below the continuum thresholds. Finally, we perform the double Borel transform to obtain the QCD sum rules.

In Ref.\cite{WZG-ZJX-Zc-Decay},  we suggested a novel approach to calculate three-hadron coupling constants  with  the three-point QCD sum rules based on the rigorous  duality for the first time,   thereafter, the rigorous  duality has been successfully applied to study the strong decays of the exotic tetraquark and pentaquark (molecular) states \cite{Decay-mole-WZG-WX,WZG-Y4660-Decay,WZG-X4140-decay,WZG-X4274-decay,WZG-Z4600-decay,WZG-Pc4312-decay-tetra}.
In this work, we take the $P_c(4312)$ as the $\bar{D}\Sigma_c$ molecular state with the isospin $I=\frac{1}{2}$ and extend our previous work to explore its two-body strong decays with the QCD sum rules based on rigorous  duality. Furthermore, we investigate the decays of its higher isospin cousin $P_c(4330)$ as a cross-check of the molecule assignment, the observation of the $P_c(4330)$ in the $J/\psi \Delta$ and $\eta_c \Delta$ decay modes would shed light on the nature of the $P_c$ states.

In our previous work, we observed that there maybe exist a resonant molecular state $P_c(4330)$ with the isospin $(I,I_3)=(\frac{3}{2},\frac{1}{2})$ as the cousin of the $P_c(4312)$ based on the QCD sum rules  \cite{mass-mole-WXW-SCPMA}. We cannot assign a hadron by the mass alone, we should explore the decays  at least.   Now we explore its two-body strong decays to  select the best channel to search for it experimentally, and try to provide a guide for the high energy physics experiments.   If the predictions  can be testified, it would  in return prove our interpretations for the nature of the $P_c(4312)$. In Refs.\cite{mass-mole-WXW-SCPMA,mass-mole-WXW-IJMPA,mass-mole-WXW-CPC}, we distinguish the isospins and investigate the hidden-charm molecular states without strange, with strange and with double-strange in a comprehensive way, it is very interesting to explore  all the two-body strong decays with the QCD sum rules (also including the mixing effects), and make predictions on the partial decay widths to provide a valuable guide for the international high energy physics experiments, however, we can only accomplish the tedious calculations one by one.

The article is arranged as follows: we obtain the QCD sum rules for the strong decays of $P_c(4312)$ and $P_c(4330)$ in Sect.2; we obtain the QCD sum rules for the $\Delta$ baryon with isospin $(I,I_3)=(\frac{3}{2},\frac{1}{2})$ in Sect.3; we present the numerical results and discussions in Sect.4; and Sect.5 is reserved for our conclusions.

\section{QCD sum rules for  strong decays of the pentaquark molecular states}
Followed Ref.\cite{mass-mole-WXW-SCPMA}, the quantum numbers $(I,J^P)$ of the $P_c(4312)$ and $P_c(4330)$ are $(\frac{1}{2},\frac{1}{2}^-)$ and $(\frac{3}{2},\frac{1}{2}^-)$, respectively. Considering  conservation of the isospin $I$ in the strong decays, we study the following processes,
\begin{eqnarray}
P_c(4312)&\rightarrow&\eta_c+N\, , \nonumber\\
P_c(4312)&\rightarrow&J/\psi+N\, , \nonumber\\
P_c(4330)&\rightarrow&\eta_c+\Delta\, , \nonumber\\
P_c(4330)&\rightarrow&J/\psi+\Delta\, ,
\end{eqnarray}
where the $N$ represents the proton to avoid confusion due to the four momentum $p_\mu$, and the $\Delta$ baryon has the $(I,J^P)=(\frac{3}{2},\frac{3}{2}^+)$. We apply the currents $J_{\eta_c}(x)$, $J_{J/\psi,\mu}(x)$, $J_{N}(x)$, $J_{\Delta,\mu}(x)$, $J_{P}(x)$ and $J_{P'}(x)$ to  interpolate the $\eta_c$, $J/\psi$, $N$, $\Delta$, $P_c(4312)$ and $P_c(4330)$, respectively, and list out them explicitly,
\begin{eqnarray}\label{Currents-meson}
J_{\eta_c}(x) &=& \overline{c}(x)i\gamma_5c(x)\, , \nonumber\\
J_{J/\psi,\mu}(x) &=& \overline{c}(x)\gamma_\mu c(x)\, , \nonumber\\
J_{N}(x)      &=& \varepsilon^{ijk}u^{iT}(x)C\gamma_{\alpha}u^{j}(x)\gamma^{\alpha}\gamma_5d^{k}(x)\, , \nonumber\\
J_{\Delta,\mu}(x) &=& \frac{1}{\sqrt{3}} \varepsilon^{ijk}u^{iT}(x)C\gamma_{\mu}u^{j}(x)d^{k}(x)+\sqrt{\frac{2}{3}} \varepsilon^{ijk}u^{iT}(x)C\gamma_{\mu}d^{j}(x)u^{k}(x)\, ,
\end{eqnarray}
\begin{eqnarray}\label{Currents-mole}
J_{P}(x) &=& \frac{1}{\sqrt{3}}\varepsilon^{ijk}\overline{c}(x)i\gamma_5u(x)u^{iT}(x)C\gamma_{\alpha}d^{j}(x)\gamma^{\alpha}\gamma_5c^{k}(x)\nonumber\\
&& -\sqrt{\frac{2}{3}}\varepsilon^{ijk}\overline{c}(x)i\gamma_5d(x)u^{iT}(x)C\gamma_{\alpha}u^{j}(x)\gamma^{\alpha}\gamma_5c^{k}(x)\, , \nonumber\\
J_{P'}(x) &=& \sqrt{\frac{2}{3}}\varepsilon^{ijk}\overline{c}(x)i\gamma_5u(x)u^{iT}(x)C\gamma_{\alpha}d^{j}(x)\gamma^{\alpha}\gamma_5c^{k}(x)\nonumber\\
&&+\frac{1}{\sqrt{3}}\varepsilon^{ijk}\overline{c}(x)i\gamma_5d(x)u^{iT}(x)C\gamma_{\alpha}u^{j}(x)\gamma^{\alpha}\gamma_5c^{k}(x)\, ,
\end{eqnarray}
where the $C$ is the charge conjugation matrix, and the $i$, $j$ and $k$ are the color indexes.
We can observe  that the interpolating currents, see Eq.\eqref{Currents-mole}, for the lower and higher isospins of the $P_c$ states, are different from the one constructed in Ref.\cite{Decay-mole-WZG-WX}, where we do not consider clear definition of the isospin. Thus, the calculations  of all the related hadronic  coupling constants are of great difference at the QCD sides.

In general, we can construct several currents to interpolate one hadron or construct one current to interpolate several hadrons, as a hadron maybe have several Fock components. The currents with the same quantum numbers could mix with each other under renormalization, we should introduce the mixing matrixes $U$ to obtain the diagonal currents, $J^{\prime }=UJ$, which couple potentially to (more) physical states. The matrixes $U$ can be determined by direct calculating anomalous dimensions of the current operators, however, up to today, even for the conventional baryons,  we cannot obtain a diagonal current, which is an special superposition of several non-trial currents to match with all the considerable Fock states. In the present work, we only consider the $\bar{D}\Sigma_c$ components of the $P_c(4312)$ and $P_c(4330)$ with definite spins and isospins. In fact, the $P_c(4312)$ and $P_c(4330)$ maybe have other important Fock components, such as the $\eta_c N$, $\bar{D}^*\Sigma_c$, $\bar{D}^*\Sigma_c^*$, $\eta_c \Delta$, etc, where the $\bar{D}$, $\bar{D}^*$, $\cdots$, $\Sigma_c$, $\Sigma_c^*$, $\cdots$ denote the color-singlet clusters having the same quantum numbers as the conventional mesons and baryons, not the physical mesons and baryons, as we choose the local five-quark currents.      We prefer explore the possibility and accomplish the tedious task in our next work, as it will take several months at least. In the present work, we calculate the correlation functions with the full QCD, while in Ref.\cite{Decay-mole-Sakai}, only heavy quark symmetry and phenomenological contact four-hadron interactions are retained, we should not be surprised if different conclusions are obtained, all the predictions should be confronted to the experimental data in the future.

Now we write down the three-point correlation functions in the QCD sum rules,
\begin{eqnarray}\label{CF-Pi}
\Pi(p,q) &=& i^2 \int d^4xd^4y e^{ip\cdot x}e^{iq\cdot y} \langle 0|T \left\{ J_{\eta_c}(x)J_N(y) \bar{J}_{P}(0) \right\}|0\rangle\, ,
\end{eqnarray}
\begin{eqnarray}\label{CF-Pi-mu-1}
\Pi_{1,\mu}(p,q) &=& i^2 \int d^4xd^4y e^{ip\cdot x}e^{iq\cdot y} \langle 0|T \left\{ J_{J/\psi,\mu}(x)J_N(y) \bar{J}_{P}(0) \right\}|0\rangle\, ,
\end{eqnarray}
\begin{eqnarray}\label{CF-Pi-mu-2}
\Pi_{2,\mu}(p,q) &=& i^2 \int d^4xd^4y e^{ip\cdot x}e^{iq\cdot y} \langle 0|T \left\{ J_{\eta_c}(x)J_{\Delta,\mu}(y) \bar{J}_{P'}(0) \right\}|0\rangle\, ,
\end{eqnarray}
\begin{eqnarray}\label{CF-Pi-mu-nu}
\Pi_{\mu\nu}(p,q) &=& i^2 \int d^4xd^4y e^{ip\cdot x}e^{iq\cdot y} \langle 0|T \left\{ J_{J/\psi,\mu}(x)J_{\Delta,\nu}(y) \bar{J}_{P'}(0) \right\}|0\rangle\, .
\end{eqnarray}

At the hadron side, we insert the complete sets of intermediate hadron states with the same quantum numbers as the currents
$J_{\eta_c}(x)$, $J_{J/\psi,\mu}(x)$, $J_{N}(x)$, $J_{\Delta,\mu}(x)$, $J_{P}(x)$ and $J_{P'}(x)$ into those three-point correlation functions and isolate the contributions  of the ground states,
\begin{eqnarray}
\Pi(p,q) &=& \frac{f_{\eta_c}m_{\eta_c}^2}{2m_c}\lambda_N\lambda_{P} g_{P} \frac{\left(\!\not\!{q}+m_N\right)\left(\!\not\!{p'}+m_{P}\right)}
{\left(m^2_{P}-p'^2\right)\left(m^2_{\eta_c}-p^2\right)\left(m^2_{N}-q^2\right)}+\cdots\, ,
\end{eqnarray}
\begin{eqnarray}
\Pi_{1,\mu}(p,q)&=&f_{J/\psi}m_{J/\psi}\lambda_N\lambda_{P}\frac{-i}{\left(m^2_{P}-p'^2\right)\left(m^2_{J/\psi}-p^2\right)\left(m^2_{N}-q^2\right)}\left( -g_{\mu\alpha}+\frac{p_{\alpha}p_{\mu}}{p^2}\right)\nonumber\\
&&\left(\!\not\!{q}+m_N\right)\left(g_V\gamma^{\alpha}-\frac{ig_T}{m_{P}+m_N}\sigma^{\alpha\beta}p_{\beta}\right)
\gamma_5\left(\!\not\!{p'}+m_{P}\right)+\cdots\, ,
\end{eqnarray}
\begin{eqnarray}
\Pi_{2,\mu}(p,q) &=& \frac{f_{\eta_c}m_{\eta_c}^2}{2m_c}\frac{g_{P'}\lambda_\Delta\lambda_{P'}p^{\alpha}}
{\left(m^2_{P'}-p'^2\right)\left(m^2_{\eta_c}-p^2\right)\left(m^2_{\Delta}-q^2\right)}\nonumber\\
&& \left(\!\not\!{q}+m_{\Delta}\right)\left( g_{\mu\alpha}-\frac{1}{3}\gamma_\mu\gamma_\alpha-\frac{2}{3}\frac{q_\mu q_\alpha}{m_{\Delta}^2}+\frac{q_\mu\gamma_\alpha-q_{\alpha}\gamma_\mu}{3m_\Delta} \right)\gamma_5\left(\!\not\!{p'}+m_{P'}\right)+\cdots\,,
\end{eqnarray}
\begin{eqnarray}
\Pi_{\mu\nu}(p,q)&=& f_{J/\psi}m_{J/\psi}\lambda_{\Delta}\lambda_{P'}\frac{-1}{\left(m^2_{P'}-p'^2\right)\left(m^2_{J/\psi}-p^2\right)\left(m^2_{\Delta}-q^2\right)}
\nonumber  \\
&&\left(\!\not\!{q}+m_{\Delta}\right)\left( g_{\nu\alpha}-\frac{1}{3}\gamma_\nu\gamma_\alpha-\frac{2}{3}\frac{q_\nu q_\alpha}{m_{\Delta}^2}+\frac{q_\nu\gamma_\alpha-q_{\alpha}\gamma_\nu}{3m_\Delta} \right)\varepsilon_{\mu}\nonumber  \\
&&  \left[ g_A(p_{\alpha}\!\not\!{\varepsilon}-\varepsilon_{\alpha}\!\not\!{p})-g_B(\verb"P"\cdot\varepsilon p_{\alpha}-\verb"P"\cdot p\varepsilon_{\alpha})+g_C(-p^2\varepsilon_{\alpha}) \right]\left(\!\not\!{p'}+m_{P'}\right)+\cdots\,,
\end{eqnarray}
where the $g_{P}$, $g_{P'}$, $g_V$, $g_T$, $g_A$, $g_B$ and $g_C$ are the hadronic coupling constants, the $\lambda_{P}$, $\lambda_{P'}$, $\lambda_{N}$ and $\lambda_{\Delta}$ are the pole residues, the $f_{\eta_c}$ and $f_{J/\psi}$ are the decay constants, $\verb"P"=\frac{q+p'}{2}$, and we have applied the definitions,
\begin{eqnarray}
\langle 0| J_{P}(0)|\mathcal{P}_{P}(p')\rangle &=& \lambda_{P}U_{P}(p')\,,\nonumber  \\
\langle 0| J_{P'}(0)|\mathcal{P}_{P'}(p')\rangle &=& \lambda_{P'}U_{P'}(p')\,,\nonumber  \\
\langle 0| J_{N}(0)|N(q)\rangle &=& \lambda_NU(q)\,,\nonumber \\
\langle 0| J_{\Delta,\mu}(0)|\Delta(q)\rangle &=& \lambda_{\Delta}U_{\mu}(q)\,,\nonumber \\
\langle 0| J_{J/\psi,\mu}(0)|J/\psi(p)\rangle &=& f_{J/\psi}m_{J/\psi}\varepsilon_{\mu}\,,\nonumber \\
\langle 0| J_{\eta_c}(0)|\eta_{c}(p)\rangle &=& \frac{f_{\eta_c}m_{\eta_c}^2}{2m_c}\,,
\end{eqnarray}
\begin{eqnarray}\label{gP-gPprime}
\langle \eta_c(p)N(q)|\mathcal{P}_{P}(p')\rangle &=& ig_{P}\overline{U}(q)U_{P}(p')\,,\nonumber \\
\langle J/\psi(p)N(q)|\mathcal{P}_{P}(p')\rangle &=& \overline{U}(q)\varepsilon_{\alpha}^*\left(g_V\gamma^{\alpha}-i\frac{g_T}{m_{P}+m_N}
\sigma^{\alpha\beta}p_{\beta}\right)\gamma_5U_{P}(p')\,,\nonumber \\
\langle \eta_c(p)\Delta(q)|\mathcal{P}_{P'}(p')\rangle &=& -ig_{P'}\overline{U}_{\alpha}(q)\gamma_5U_{P'}(p')p^{\alpha}\,,\nonumber\\
\langle J/\psi(p)\Delta(q)|\mathcal{P}_{P'}(p')\rangle &=& i\overline{U}_{\alpha}(q)[g_A(p_{\alpha}\!\not\!{\varepsilon}-\varepsilon_{\alpha}\!\not\!{p})
\nonumber\\
&&-g_B(\verb"P"\cdot \varepsilon p_{\alpha}-\verb"P"\cdot p\varepsilon_{\alpha})+g_{C}(p\cdot\varepsilon p_{\alpha}-p^2\varepsilon_{\alpha})]U_{P'}(p')\,,
\end{eqnarray}
where the  $U(q)$, $U_{P}(p')$ and $U_{P'}(p')$ are the Dirac spinors, the $U_{\mu}(q)$ is the Rarita-Schwinger spinor, the $\varepsilon_\mu$ represents the polarization vector of the $J/\psi$, $p\cdot \varepsilon=0$. The $|\eta_c\rangle$, $|J/\psi\rangle$, $|N\rangle$, $|\Delta\rangle$, $|\mathcal{P}_{P}\rangle$ and $|\mathcal{P}_{P'}\rangle$ denote the ground states of the $\eta_c$, $J/\psi$, $N$, $\Delta$, $P_{c}(4312)$ and $P_{c}(4330)$, respectively. For the definition of the vertex $\langle J/\psi(p)\Delta(q)|\mathcal{P}_{P'}(p')\rangle$, one can consult Ref.\cite{Jones-Scadron-Vertice}.

It is natural to consider that the correlation functions at the hadronic side should match that of the QCD side, that is, $\Pi_{H}(p,q)=\Pi_{ QCD}(p,q)$, where the $\Pi$ represent the correlation functions in Eqs.\eqref{CF-Pi}-\eqref{CF-Pi-mu-nu}, the subscripts $H$ and $ QCD$ denote the hadron and QCD sides, respectively. It is then reasonable to have relation $Tr[\Pi_{H}(p,q)\cdot\Gamma]=Tr[\Pi_{ QCD}(p,q)\cdot\Gamma]$, where the $\Gamma$ is the some chosen $\gamma-$matrix in the Dirac spinor space.
For the $\Pi(p,q)$ in Eq.\eqref{CF-Pi}, we choose $\Gamma=\sigma_{\mu\nu}$ and $\gamma_\mu$, and select the corresponding tensor structures $p_{\mu}q_{\nu}-q_{\mu}p_{\nu}$ and $q_\mu$, respectively. For the $\Pi_{1,\mu}(p,q)$ and $\Pi_{2,\mu}(p,q)$ in Eq.\eqref{CF-Pi-mu-1} and Eq.\eqref{CF-Pi-mu-2}, we set $\Gamma=\gamma_5\!\not\!{z}$ and $\gamma_5$, and choose the tensor structures $q_\mu p\cdot z$ and $q_\mu$, respectively \cite{Decay-mole-WZG-WX,WZG-Pc4312-decay-tetra}, where $z$ is a four vector. For the $\Pi_{\mu\nu}(p,q)$ in Eq.\eqref{CF-Pi-mu-nu}, we pick out its tensor structures $\!\not\!{p}\!\not\!{q}g_{\mu\nu}$ and $\!\not\!{p}(\gamma_{\mu}p_\nu+\gamma_{\nu}p_{\mu}+\gamma_{\mu}q_\nu+\gamma_{\nu}q_{\mu})$, simplify  $\Pi_{\mu\nu}\cdot\!\not\!{z}$ and choose the structure $\!\not\!{p}\!\not\!{q} (\gamma_{\mu}p_\nu+\gamma_{\nu}p_{\mu}+\gamma_{\mu}q_\nu+\gamma_{\nu}q_{\mu})p\cdot z$. In details, the selected structures are expressed as,
\begin{eqnarray}
\frac{1}{4}Tr[\Pi(p,q)\sigma_{\mu\nu}] &=& \Pi_a(p'^2,p^2,q^2)i(p_\mu q_\nu-q_\mu p_\nu)+\cdots\,,\nonumber \\
\frac{1}{4}Tr[\Pi(p,q)i\gamma_{\mu}] &=& \Pi_b(p'^2,p^2,q^2)iq_\mu+\cdots\,, \nonumber \\
\frac{1}{4}Tr[\Pi_{1,\mu}(p,q)\gamma_{5}\!\not\!{z}] &=& \Pi_c(p'^2,p^2,q^2)iq_\mu p\cdot z+\cdots\,, \nonumber\\
\frac{1}{4}Tr[\Pi_{1,\mu}(p,q)\gamma_{5}] &=& \Pi_d(p'^2,p^2,q^2)iq_\mu +\cdots\,,\nonumber \\
\frac{1}{4}Tr[\Pi_{2,\mu}(p,q)\gamma_{5}\!\not\!{z}] &=& \Pi_e(p'^2,p^2,q^2)iq_\mu p\cdot z+\cdots\,, \nonumber\\
\frac{1}{4}Tr[\Pi_{2,\mu}(p,q)\gamma_{5}] &=& \Pi_f(p'^2,p^2,q^2)iq_\mu +\cdots\,,
\end{eqnarray}
\begin{eqnarray}
\Pi_{\mu\nu}(p,q)  &=& \Pi_A(p'^2,p^2,q^2)\!\not\!{p}\!\not\!{q}g_{\mu\nu}+ \Pi_C(p'^2,p^2,q^2)\!\not\!{p} (\gamma_{\mu}p_\nu+\gamma_{\nu}p_{\mu}+\gamma_{\mu}q_\nu+\gamma_{\nu}q_{\mu})+\cdots\,,\nonumber \\
\Pi_{\mu\nu}(p,q)\cdot \!\not\!{z}  &=& \Pi_B(p'^2,p^2,q^2)\!\not\!{p}\!\not\!{q} (\gamma_{\mu}p_\nu+\gamma_{\nu}p_{\mu}+\gamma_{\mu}q_\nu+\gamma_{\nu}q_{\mu})p\cdot z+\cdots\, .
\end{eqnarray}
We apply the same analysis for the components of the correlation functions $\Pi_Z(p'^2,p^2,q^2)$ as discussed in Refs.\cite{Decay-mole-WZG-WX,WZG-Pc4312-decay-tetra}, where the $Z$ represents  $a$, $b$, $\cdots$, $f$, $A$, $B$ and $C$.

At the QCD side, after accomplishing the operator product expansion, we perform the trace and choose the selected tensor structures. For the relation among $p'$, $p$ and $q$, they satisfy $p'=p+q$ for all the decays. We set $p'^2=\xi p^2$, where the $\xi$ is a  parameter. For example, in the decay $P_c(4312)\rightarrow\eta_c+N$, $0\leq \xi\leq \frac{2m_N^2}{m_{\eta_c^2}}+2$, we can set $\xi= \frac{m_N^2}{m_{\eta_c^2}}+1$.
Just like in our previous works \cite{Decay-mole-WZG-WX,WZG-ZJX-Zc-Decay,WZG-Y4660-Decay,WZG-X4140-decay,WZG-X4274-decay,WZG-Z4600-decay,WZG-Pc4312-decay-tetra}, we take rigorous quark-hadron duality below the continuum thresholds, and perform double Borel transformation and obtain the QCD sum rules for the hadronic coupling constants,
\begin{eqnarray}\label{QCDSG-G-i}
&& \frac{f_{\eta_c}m_{\eta_c}^2\lambda_N\lambda_{P} g_{P}^a}{2m_c \xi}\frac{1}{\frac{m_{P}^2}{\xi}-m_{\eta_c}^2}\left\{ {\rm exp} \left( -\frac{m_{\eta_c}^2}{T_1^2} \right)-{\rm exp} \left( -\frac{m_{P}^2}{\xi T_1^2} \right) \right\}{\rm exp} \left( -\frac{m_N^2}{T_2^2} \right) \nonumber \\
&& +C_a {\rm exp} \left( -\frac{m_{\eta_c}^2}{T_1^2}-\frac{m_N^2}{T_2^2} \right)=\int_{4m_c^2}^{s_{\eta_c}^0}ds\int_0^{s_N^0}du\, \rho_a(s,u){\rm exp}\left( -\frac{s}{T_1^2}-\frac{u}{T_2^2} \right)\,,
\end{eqnarray}

\begin{eqnarray}
&& \frac{f_{\eta_c}m_{\eta_c}^2\lambda_N\lambda_{P} g_{P}^b}{2m_c \xi}\frac{m_{P}+m_N}{\frac{m_{P}^2}{\xi}-m_{\eta_c}^2}\left\{ {\rm exp} \left( -\frac{m_{\eta_c}^2}{T_1^2} \right)-{\rm exp} \left( -\frac{m_{P}^2}{\xi T_1^2} \right) \right\}{\rm exp} \left( -\frac{m_N^2}{T_2^2} \right)\nonumber \\
&& +C_b {\rm exp} \left( -\frac{m_{\eta_c}^2}{T_1^2}-\frac{m_N^2}{T_2^2} \right)=\int_{4m_c^2}^{s_{\eta_c}^0}ds\int_0^{s_N^0}du\, \rho_b(s,u){\rm exp}\left( -\frac{s}{T_1^2}-\frac{u}{T_2^2} \right)\,,
\end{eqnarray}
\begin{eqnarray}
&& \frac{f_{J/\psi}m_{J/\psi}\lambda_N\lambda_{P} }{\xi}\frac{g_{T/V}}{\frac{m_{P}^2}{\xi}-m_{J/\psi}^2}\left\{ {\rm exp} \left( -\frac{m_{J/\psi}^2}{T_1^2} \right)-{\rm exp} \left( -\frac{m_{P}^2}{\xi T_1^2} \right) \right\}{\rm exp} \left( -\frac{m_N^2}{T_2^2} \right) \nonumber \\
&& +C_{T/V} {\rm exp} \left( -\frac{m_{J/\psi}^2}{T_1^2}-\frac{m_N^2}{T_2^2} \right)=\int_{4m_c^2}^{s_{J/\psi}^0}ds\int_0^{s_N^0}du\, \rho_{T/V}(s,u){\rm exp}\left( -\frac{s}{T_1^2}-\frac{u}{T_2^2} \right)\,,
\end{eqnarray}

\begin{eqnarray}
&& \frac{f_{\eta_c}m_{\eta_c}^2\lambda_{\Delta}\lambda_{P'} g_{P'}^e}{4m_cm_{\Delta}\xi}\frac{M_X^2}{\frac{m_{P'}^2}{\xi}-m_{\eta_c}^2}\left\{ {\rm exp} \left( -\frac{m_{\eta_c}^2}{T_1^2} \right)-{\rm exp} \left( -\frac{m_{P'}^2}{\xi T_1^2} \right) \right\}{\rm exp} \left( -\frac{m_{\Delta}^2}{T_2^2} \right)\nonumber \\
&& +C_{e} {\rm exp} \left( -\frac{m_{\eta_c}^2}{T_1^2}-\frac{m_{\Delta}^2}{T_2^2} \right)=\int_{4m_c^2}^{s_{\eta_c}^0}ds\int_0^{s_{\Delta}^0}du\, \rho_{e}(s,u){\rm exp}\left( -\frac{s}{T_1^2}-\frac{u}{T_2^2} \right)\,,
\end{eqnarray}

\begin{eqnarray}
&& \frac{f_{\eta_c}m_{\eta_c}^2\lambda_{\Delta}\lambda_{P'} g_{P'}^f }{12m_cm_{\Delta}^2\xi}\frac{M_X^2\left( 2m_{\Delta}m_{P'}-M_X^2-2m_{\Delta}^2 \right)}{\frac{m_{P'}^2}{\xi}-m_{\eta_c}^2}\left\{ {\rm exp} \left( -\frac{m_{\eta_c}^2}{T_1^2} \right)-{\rm exp} \left( -\frac{m_{P'}^2}{\xi T_1^2} \right) \right\}{\rm exp} \left( -\frac{m_{\Delta}^2}{T_2^2} \right)\nonumber \\
&& +C_{f} {\rm exp} \left( -\frac{m_{\eta_c}^2}{T_1^2}-\frac{m_{\Delta}^2}{T_2^2} \right)=\int_{4m_c^2}^{s_{\eta_c}^0}ds\int_0^{s_{\Delta}^0}du\, \rho_{f}(s,u){\rm exp}\left( -\frac{s}{T_1^2}-\frac{u}{T_2^2} \right)\,,
\end{eqnarray}

\begin{eqnarray}\label{QCDSG-G-f}
&& \frac{f_{J/\psi}m_{J/\psi}\lambda_{\Delta}\lambda_{P'} }{\xi}\frac{1}{\frac{m_{P'}^2}{\xi}-m_{J/\psi}^2}\left\{ {\rm exp} \left( -\frac{m_{J/\psi}^2}{T_1^2} \right)-{\rm exp} \left( -\frac{m_{P'}^2}{\xi T_1^2} \right) \right\}{\rm exp} \left( -\frac{m_{\Delta}^2}{T_2^2} \right)\nonumber \\
&&  K\cdot\begin{bmatrix}g_A\\g_B\\g_C\end{bmatrix}+\begin{bmatrix}C_A\\C_B\\C_C\end{bmatrix} {\rm exp} \left( -\frac{m_{J/\psi}^2}{T_1^2}-\frac{m_{\Delta}^2}{T_2^2} \right) \nonumber \\
&&=\int_{4m_c^2}^{s_{J/\psi}^0}ds\int_0^{s_{\Delta}^0}du
\begin{bmatrix} \rho_{A}(s,u)\\ \rho_{B}(s,u)\\ \rho_{C}(s,u) \end{bmatrix}{\rm exp}\left( -\frac{s}{T_1^2}-\frac{u}{T_2^2} \right)\,,
\end{eqnarray}
where
\begin{eqnarray}
M_X^2 &=& m_{P'}^2-m_{\Delta}^2-m_{\eta_c}^2 \,,\nonumber \\
M_Y^2 &=& m_{\Delta}^2+2m_{\Delta}m_{P'}+m_{P'}^2-m_{J/\psi}^2 \,,\nonumber\\
C_T &=& \left[ (m_{P}-m_N)C_c+C_d \right]\frac{m_{P}+m_N}{m_{P}^2-m_N^2-m_{J/\psi}^2}\,,\nonumber\\
\rho_T(s,u) &=& \left[ (m_{P}-m_N)\rho_c(s,u)+\rho_d(s,u) \right]\frac{m_{P}+m_N}{m_{P}^2-m_N^2-m_{J/\psi}^2}\,,\nonumber\\
C_V &=& \left( \frac{m_{J/\psi}^2}{m_{P}+m_N}C_c+C_d \right)\frac{m_{P}+m_N}{m_{P}^2-m_N^2-m_{J/\psi}^2}\,,\nonumber\\
\rho_V(s,u) &=& \left( \frac{m_{J/\psi}^2}{m_P+m_N}\rho_c(s,u)+\rho_d(s,u) \right)\frac{m_{P}+m_N}{m_{P}^2-m_N^2-m_{J/\psi}^2}\,,\nonumber\\
K &=& \begin{bmatrix} \frac{M_Y^2}{3m_{\Delta}}\left(4m_{\Delta}^2 -M_Y^2\right)&\frac{2}{3}M_Y^2(m_{P'}^2-m_{\Delta}^2)&-\frac{4}{3}m_{J/\psi}^2M_Y^2
\\ \frac{4(m_{\Delta}m_{P'}-m_{P'}^2+m_{J/\psi}^2)}{3m_{\Delta}^2}&
-\frac{2(m_{\Delta}^2-m_{P'}^2+3m_{J/\psi}^2)}{3m_{\Delta}}&\frac{8m_{J/\psi}^2}{3m_{\Delta}}\\
\frac{4}{3}m_{\Delta}&-\frac{2}{3}M_Y^2&\frac{4}{3}M_Y^2\nonumber
\end{bmatrix}\,,
\end{eqnarray}
and
 \begin{eqnarray}
\rho_{Z}(s,u)&=& {\lim_{\epsilon_2\to 0}} \,\,{\lim_{\epsilon_1\to 0}}\,\,\frac{  {\rm Im}_{s}\,{\rm Im}_{u}\,\Pi_{Z}(p^{\prime 2},s+i\epsilon_2,u+i\epsilon_1) }{\pi^2} \, ,
\end{eqnarray}
  are the spectral densities at the QCD sides, we add the superscripts $a$ and $b$ to denote the hadronic coupling constant $g_{P}$ from the components $\Pi_a(p'^2,p^2,q^2)$ and $\Pi_b(p'^2,p^2,q^2)$, respectively, and add the superscripts $e$ and $f$ to denote the hadronic coupling constant $g_{P'}$ from the components $\Pi_e(p'^2,p^2,q^2)$ and $\Pi_f(p'^2,p^2,q^2)$, respectively; the $C_Z$ denote the unknown parameters, which are determined in the numerical calculations to obtain flat platforms \cite{Decay-mole-WZG-WX,WZG-ZJX-Zc-Decay,WZG-Y4660-Decay,WZG-X4140-decay,
  WZG-X4274-decay,WZG-Z4600-decay,WZG-Pc4312-decay-tetra}.

\section{QCD sum rules for the $\Delta$ baryon with isospin $I=\frac{3}{2}$}
In Eq.\eqref{Currents-meson}, the current $J_{\Delta,\mu}(x)$ is the isospin eigenstate $|II_3\rangle=|\frac{3}{2}\frac{1}{2}\rangle$. We write down the two-point correlation function,
\begin{eqnarray}
\Pi_{\Delta,\mu\nu}(q)&=&i\int d^4y e^{iq\cdot y} \langle 0|T\left\{ J_{\Delta,\mu}(y)\bar{J}_{\Delta,\nu}(0)\right\}|0\rangle\, .
\end{eqnarray}
At the hadron side, we insert a complete set of baryon states with the same quantum number as the current into the correlation function $\Pi_{\Delta,\mu\nu}$ and isolate the contribution of the ground state,
\begin{eqnarray}
\Pi_{\Delta,\mu\nu}(q) &=&\lambda_\Delta^2 \frac{\!\not\!{q}+m_\Delta}{m_\Delta^2-q^2}\left(-g_{\mu\nu}+\cdots\right)+\cdots\, .
\end{eqnarray}
We choose tensor structures $\!\not\!{q}g_{\mu\nu}$ and $g_{\mu\nu}$ for analysis. After accomplishing the operator product expansion, we obtain the corresponding  spectral densities $\rho^1_{QCD}(u)$ and $\rho^0_{QCD}(u)$, respectively, then adopt quark-hadron duality, and obtain two QCD sum rules,
\begin{eqnarray}\label{SR-Delta}
\lambda_{\Delta}^2 {\exp}\left(-\frac{m_{\Delta}^2}{T^2}\right)&=& \int_0^{s_{\Delta}^0}\rho^1_{QCD}(u)\,{\rm exp} \left(-\frac{u}{T^2}\right)du\,,\nonumber\\
m_{\Delta}\lambda_{\Delta}^2 {\exp}\left(-\frac{m_{\Delta}^2}{T^2}\right)&=& \int_0^{s_{\Delta}^0}\rho^0_{QCD}(u)\,{\rm exp} \left(-\frac{u}{T^2}\right)du\, ,
\end{eqnarray}
where
\begin{eqnarray}
\rho^1_{QCD}(u) &=& \left(\frac{\sqrt{2}}{2560} + \frac{7}{3072}\right)   \frac{u^2}{\pi^{4}} - \left(\frac{17}{13824 \sqrt{2}} + \frac{49}{55296}\right) \langle g_s^2GG\rangle   \frac{1}{ \pi^{4}}\nonumber\\
 &&+ \left(\frac{1}{3 \sqrt{2}} + \frac{11}{36}\right) \langle\bar{q}q\rangle^2  \delta\left(u\right) +  \frac{ g_s^2 \langle\bar{q}q\rangle^2}{324}   \frac{\delta(u) }{\pi^{2}}\nonumber\\
&& - \left(\frac{23}{108 \sqrt{2}} + \frac{13}{54}\right) \langle\bar{q}q\rangle \langle\bar{q}g_s\sigma Gq\rangle \frac{\delta(u)}{T^2}\,,
\end{eqnarray}
\begin{eqnarray}
\rho^0_{QCD}(u) &=&  \left(\frac{ \sqrt{2}}{72}-\frac{1}{8 \sqrt{2}} - \frac{23}{288}\right) \langle\bar{q}q\rangle \frac{u}{\pi^{2}} + \left(\frac{1}{16 \sqrt{2}} + \frac{7}{192}\right) \langle\bar{q}g_s\sigma Gq\rangle \frac{1}{\pi^{2}}\nonumber\\
&&+ \left(\frac{1}{288 \sqrt{2}} + \frac{11}{3456}\right) \langle g_s^2GG\rangle \langle\bar{q}q\rangle \frac{\delta(u)}{\pi^{2}}\nonumber\\
&&+ \left(\frac{4}{729 \sqrt{2}} + \frac{5}{729}\right) g_s^2\langle\bar{q}q\rangle^3  \frac{\delta(u)}{T^2}\,.
\end{eqnarray}

We differentiate  Eq.\eqref{SR-Delta} with respect to $\tau=\frac{1}{T^2}$, then obtain  the mass
  $m_\Delta$ through a fraction,
\begin{eqnarray}
m^{2}_{\Delta} &=& \frac{-\frac{\partial}{\partial \tau}\int_0^{s_{\Delta}^0}\rho^{1/0}_{QCD}(u){\rm exp}\left(-\frac{u}{T^2}\right)du}{\int_0^{s_{\Delta}^0}\rho^{1/0}_{QCD}(u){\rm exp}\left(-\frac{u}{T^2}\right)du}\,.
\end{eqnarray}

\section{Numerical results and discussions}
The standard values of the vacuum condensates are listed as $\langle\overline{q}q\rangle=-(0.24\pm0.01\,{\rm GeV})^3$, $\langle\overline{q}g_s\sigma Gq\rangle=m_0^2\langle\overline{q}q\rangle$, $m_0^2=(0.8\pm0.1)\,{\rm GeV}^2$, $\langle\frac{\alpha_s}{\pi}GG\rangle=(0.33\,{\rm GeV})^4$ at the energy scale $\mu=1\,{\rm GeV}$ \cite{SVZ1,SVZ2,Reinders,ColangeloReview}, and we apply the value of the $
\overline{MS}$ mass $m_c(m_c)=1.275\pm0.025\,{\rm GeV}$ from the Particle Data Group \cite{PDG}. The energy-scale dependence of those parameters are written as,
\begin{eqnarray}
\notag \langle\overline{q}q\rangle(\mu)&&=\langle\overline{q}q\rangle(1{\rm GeV})\left[\frac{\alpha_s(1{\rm GeV})}{\alpha_s(\mu)}\right]^{\frac{12}{33-2n_f}}\, ,\\
\notag \langle\overline{q}g_s\sigma Gq\rangle(\mu)&& =\langle\overline{q}g_s\sigma Gq\rangle(1{\rm GeV})\left[\frac{\alpha_s(1{\rm GeV})}{\alpha_s(\mu)}\right]^{\frac{2}{33-2n_f}}\, ,\\
\notag  m_c(\mu)&&=m_c(m_c)\left[\frac{\alpha_s(\mu)}{\alpha_s(m_c)}\right]^{\frac{12}{33-2n_f}}\, ,\\
\notag \alpha_s(\mu)&&=\frac{1}{b_0t}\left[1-\frac{b_1}{b_0^2}\frac{\rm{log}\emph{t}}{t}+\frac{b_1^2(\rm{log}^2\emph{t}-\rm{log}\emph {t}-1)+\emph{b}_0\emph{b}_2}{b_0^4t^2}\right]\, ,
\end{eqnarray}
where $t={\rm log}\frac{\mu^2}{\Lambda_{QCD}^2}$, $\emph b_0=\frac{33-2\emph{n}_\emph{f}}{12\pi}$, $b_1=\frac{153-19n_f}{24\pi^2}$, $b_2=\frac{2857-\frac{5033}{9}n_f+\frac{325}{27}n_f^2}{128\pi^3}$
and $\Lambda_{QCD}=213$ MeV, $296$ MeV, $339$ MeV for the flavors $n_f=5,4,3$, respectively \cite{PDG,Narison}, the flavor numbers $n_f$ for the $P_c$ decays and $\Delta$ state are $n_f=4$ and $n_f=3$, respectively. We set the energy scales for the decays $P_c(4312)\rightarrow \eta_c N$, $P_c(4330)\rightarrow \eta_c \Delta$ as   $\mu=\frac{1}{2}m_{\eta_c}$, and for the decays $P_c(4312)\rightarrow J/\psi N$, $P_c(4330)\rightarrow J/\psi \Delta$ as $\mu=\frac{1}{2}m_{J/\psi}$ \cite{Decay-mole-WZG-WX,EWZGHuang}. For the $\Delta$ baryon state, we set the energy scale $\mu=1$. We obey the experimental data and set $m_P=4.312\,{\rm GeV}$ \cite{RAaij2}, and we set $m_{P'}=4.330\,{\rm GeV}$ considering  the calculation/conclusion in Ref.\cite{mass-mole-WXW-SCPMA}. From the Particle Data Group \cite{PDG}, we take $m_{\eta_c}=2.984\,{\rm GeV}$, $m_N=0.938\,{\rm GeV}$, $m_{J/\psi}=3.097\,{\rm GeV}$, $m_{\Delta}=1.232\,{\rm GeV}$. For the decay constants, we choose $f_{J/\psi}=0.418\,{\rm GeV}$, $f_{\eta_c}=0.387\,{\rm GeV}$ \cite{Becirevic}. For the pole residues, we set $\lambda_N=3.20\times 10^{-2}\,{\rm GeV^3}$ \cite{EIoffe}, $\lambda_P=3.25\times10^{-3}\,{\rm GeV^6}$, $\lambda_{P'}=1.97\times10^{-3}\,{\rm GeV^6}$ \cite{mass-mole-WXW-SCPMA}. For the threshold parameters, we apply $\sqrt{s_{\eta_c}^0}=3.50\,{\rm GeV}$, $\sqrt{s_{N}^0}=1.30\,{\rm GeV}$, $\sqrt{s_{J/\psi}^0}=3.60\,{\rm GeV}$ \cite{Decay-mole-WZG-WX}.

\begin{figure}
 \centering
 \includegraphics[totalheight=5cm,width=7cm]{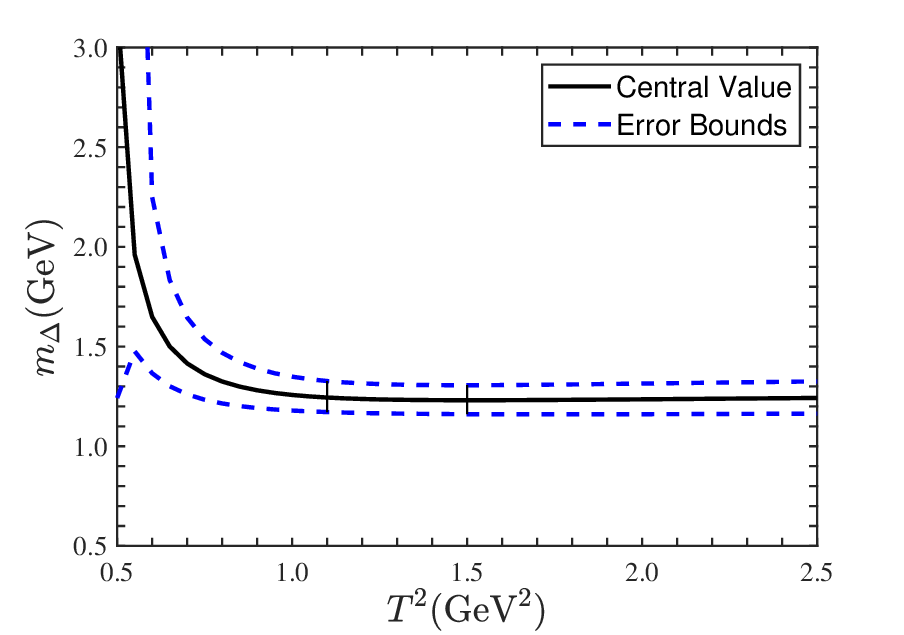}
 \includegraphics[totalheight=5cm,width=7cm]{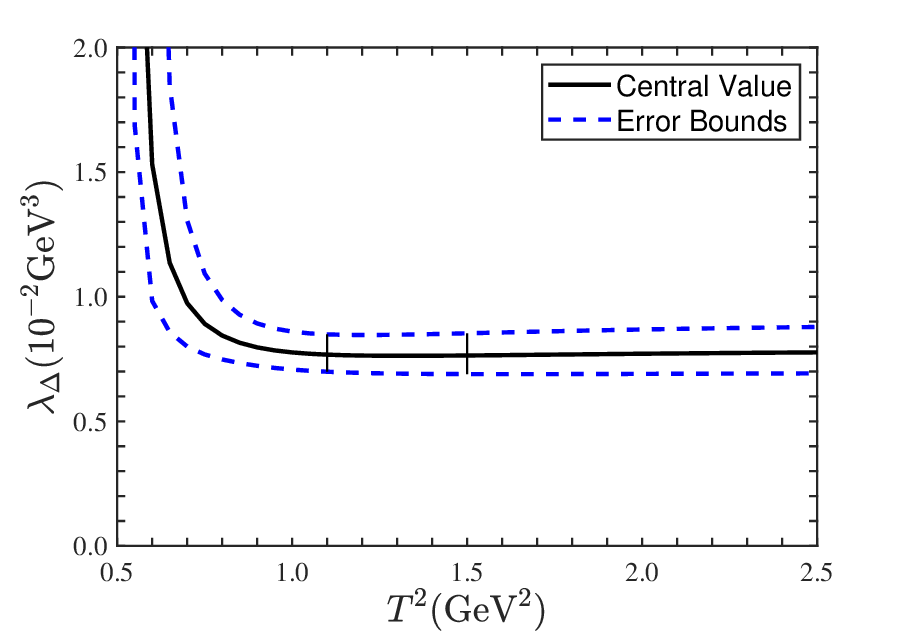}
 \caption{The $m_\Delta-T^2$ (left) and $\lambda_\Delta-T^2$ (right) curves of the $\Delta$ baryon.}\label{baryon-fig}
\end{figure}

For the $\Delta$ baryon state, we find that the numerical values of the masses in the $m_{\Delta}^0-T^2$ curve (due to $\rho_{QCD}^0(u)$) are slightly larger that in the $m_{\Delta}^1-T^2$ (due to $\rho_{QCD}^1(u)$) if taking  the same input parameters, however, it is difficult to judge which tensor structure is better, thus we phenomenologically solve the mass and pole residue of the baryon states via $m_\Delta=\frac{1}{2}(m_{\Delta}^1+m_{\Delta}^0)$ and $\lambda_\Delta=\frac{1}{2}(\lambda_{\Delta}^1+\lambda_{\Delta}^0)$, where we add the superscripts $1$ and $0$ to denote the spectral densities. The $m_{\Delta}-T^2$ and $\lambda_{\Delta}-T^2$ curves are shown in the Fig.\ref{baryon-fig}, the related parameters extracted from the Borel platforms are listed as, $\sqrt{s_{\Delta}^0}=1.61\,{\rm GeV}$, $m_{\Delta}=1.230\,{\rm GeV}$, $\lambda_{\Delta}=7.63\times 10^{-3}\,{\rm GeV^3}$, and the Borel windows $T^2=1.1-1.5\,{\rm GeV^2}$. What's more, the pole contribution in the QCD sum rules is $(41-59)\%$ and the convergency of the operator product expansion is satisfied very well.

\begin{figure}
 \centering
 \includegraphics[totalheight=5cm,width=7cm]{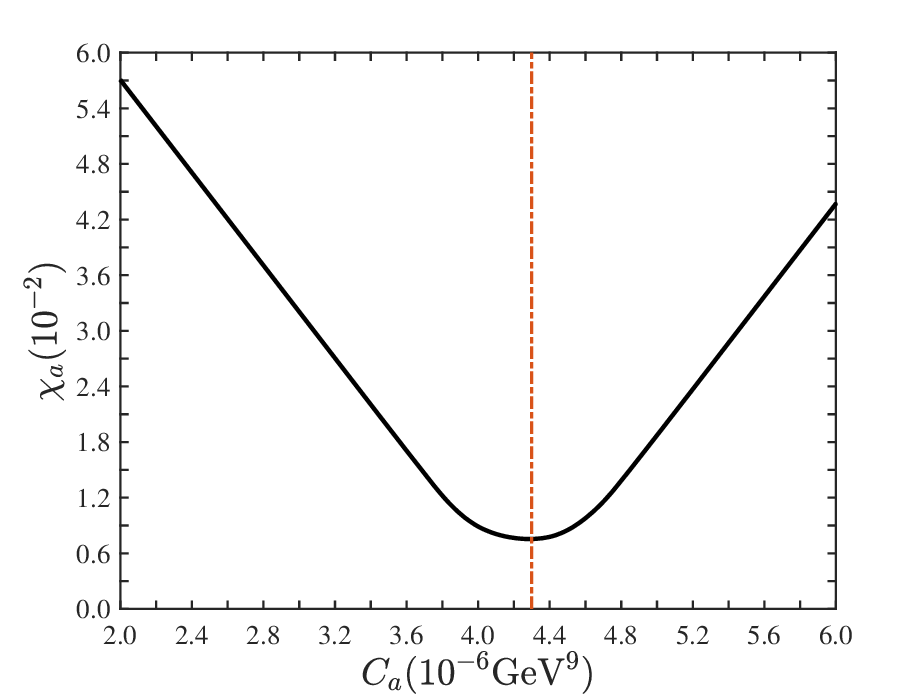}
 \caption{The free parameter $C_a$ is determined by the minimum value of $\chi_a$, where the dash-dotted line represents the chosen value of the $C_a$.}\label{Xishu}
\end{figure}

\begin{figure}
 \centering
 \includegraphics[totalheight=5cm,width=7cm]{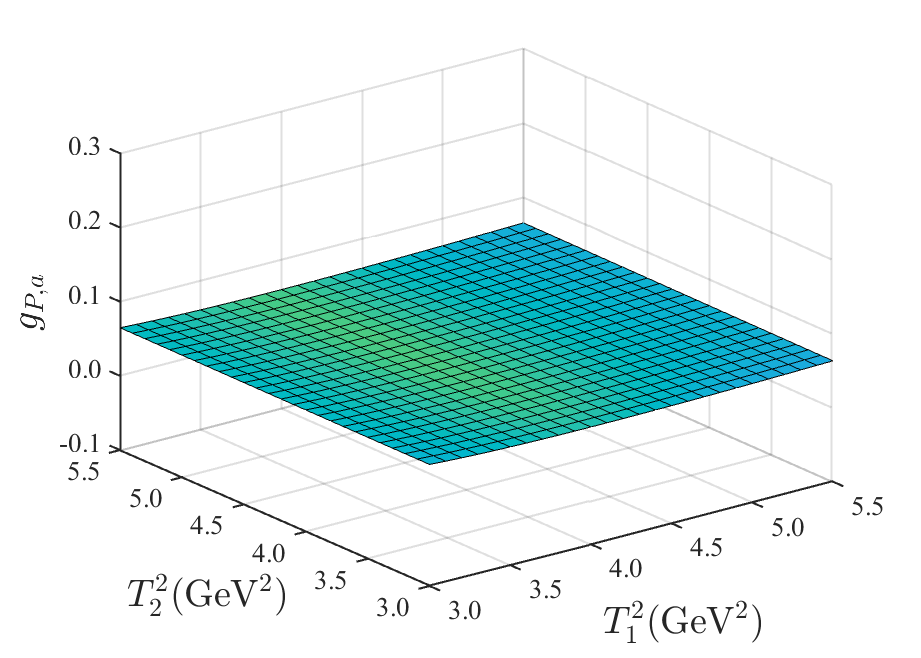}
  \includegraphics[totalheight=5cm,width=7cm]{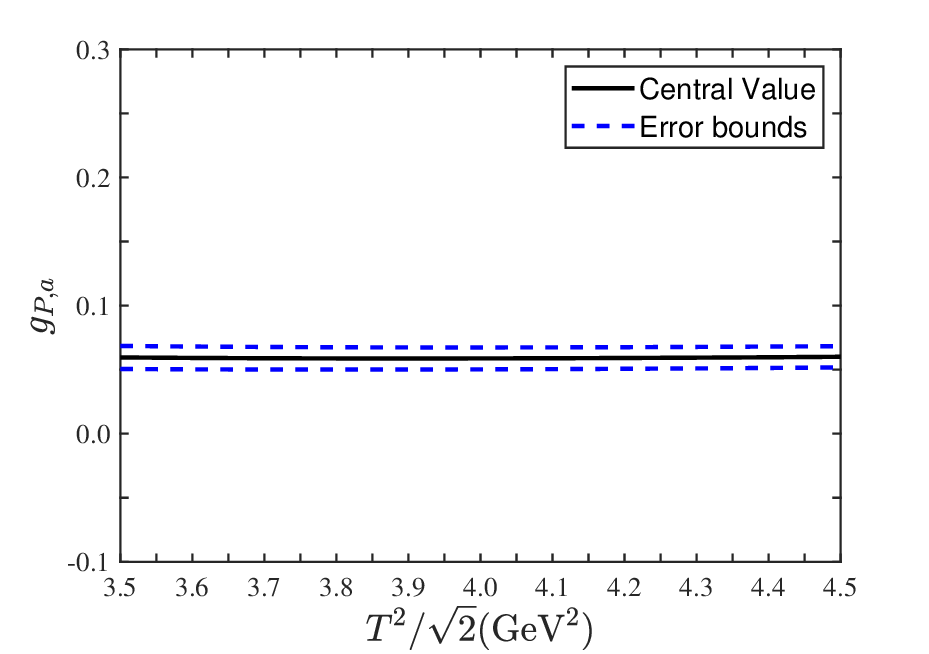}
 \caption{The $g_{P}^a(T_1^2,T_2^2)$ surface (left) and error bounds of the diagonal of the Borel platform (right), where $T^2=\sqrt{T_1^2+T_2^2}$ and $T_1^2=T_2^2$.}\label{3Dfigg1a}
\end{figure}

\begin{table}[t]
\centering
\begin{tabular}{|c|c|c|c|c|}\hline\hline
$g_Z$ & $T_1^2$ ${\rm (GeV^2)}$ & $T_2^2$ ${\rm (GeV^2)}$ & $\chi_Z$ & Values  \\ \hline
$g_{P}^a$  & $3.5-4.5$ & $3.5-4.5$ & 0.75\% &   $0.06^{+0.01}_{-0.01}$         \\ \hline
$g_{P}^b$  & $3.5-4.5$ & $3.5-4.5$ & 5.46\% &   $0.06^{+0.01}_{-0.01}$         \\ \hline
$g_{T}  $  & $5.0-6.0$ & $5.0-6.0$ & 0.83\% &   $0.36^{+0.10}_{-0.10}$         \\ \hline
$g_{V}  $  & $5.0-6.0$ & $5.0-6.0$ & 0.56\% &   $0.58^{+0.14}_{-0.12}$         \\ \hline
$g_{P'}^e$ & $4.5-5.5$ & $4.5-5.5$ & 0.79\% &   $0.30^{+0.02}_{-0.02}\,{\rm GeV}^{-1}$ \\ \hline
$g_{P'}^f$ & $4.5-5.5$ & $4.5-5.5$ & 1.59\% &   $0.30^{+0.09}_{-0.09}\,{\rm GeV}^{-1}$    \\ \hline
$g_{A}  $  & $5.5-6.5$ & $5.5-6.5$ & 2.78\% &   $0.58^{+0.08}_{-0.08}\,{\rm GeV}^{-1}$   \\ \hline
$g_{B}  $  & $6.0-7.0$ & $6.0-7.0$ & 1.74\% &   $0.43^{+0.08}_{-0.08}\,{\rm GeV}^{-2}$     \\ \hline
$g_{C}  $  & $6.0-7.0$ & $6.0-7.0$ & 2.98\% &   $0.59^{+0.08}_{-0.08}\,{\rm GeV}^{-2}$   \\ \hline
\end{tabular}
\caption{The parameters and numerical results for the hadronic coupling constants}\label{Borel-mass}
\end{table}

In the present study, the $3D$ surfaces of $g_Z=g_Z(T_1^2,T_2^2)$ are solved for the first time, where the $g_Z$ represents  the hadronic coupling constants. The Borel platforms of each coupling constants are determined under the same intervals of the Borel parameters $T_1^2$ and $T_2^2$, they are $\left(T_1^2\right)_{max}-\left(T_1^2\right)_{min}=1\,{\rm GeV^2}$ and $\left(T_2^2\right)_{max}-\left(T_2^2\right)_{min}=1\,{\rm GeV^2}$. One important requirement of the QCD sum rules is that the Borel platform should be "flat enough", which means that the error bounds originate from the Borel parameters could be neglected. For the hadronic decay constants, we argue that one can quantify the ``flat surface" via the average relative error bound $\chi_Z$, which is defined as,
\begin{eqnarray}
\chi_Z &=& \sum_{i=1}^n\frac{|g_Z(T_{1,c}^2,T_{2,c}^2)-g_Z(T_{1,i}^2,T_{2,i}^2)|}{n g_Z(T_{1,c}^2,T_{2,c}^2)}\,,
\end{eqnarray}
where the $T_{1,c}^2$ and $T_{2,c}^2$ denote the central points in the Borel platforms, the $T_{1,i}^2$ and $T_{2,i}^2$ are the selected points inside the Borel platforms, the $n$ is the number of the points on the grids in the Borel platforms. In the QCD sum rules, see Eqs.\eqref{QCDSG-G-i}-\eqref{QCDSG-G-f}, there is a free (or unknown) parameter $C_Z$ accompanying   each hadronic coupling constant $g_Z$, we  determine the optimal values of the  $C_Z$ via minimizing the $\chi_{Z}$.  Taking the $g_{P}^a$ for example, the Borel platforms are chosen via trivial and error, $\left(T_{1}^2\right)_{min}=4\,{\rm GeV^2}$, $\left(T_{2}^2\right)_{min}=4\,{\rm GeV^2}$, $\left(T_{1}^2\right)_{max}=5\,{\rm GeV^2}$ and $\left(T_{2}^2\right)_{max}=5\,{\rm GeV^2}$. The $\chi_a-C_a$ curve is shown in Fig.\ref{Xishu}, the $C_a$ is determined by the  minimum value of the $\chi_a$, i.e. $C_a=4.299\times 10^{-6}\,{\rm GeV^9}$. In an  similar way, we find $C_b=8.855 \times 10^{-5}\,{\rm GeV^{10}}$, $C_T=(-6.960{\rm GeV^2}+0.049 T_1^2-0.083T_2^2)\times 10^{-5}\,{\rm GeV^9}$, $C_V=(-7.407{\rm GeV^2}+0.053 T_1^2-0.088T_2^2)\times 10^{-5}\,{\rm GeV^9}$, $C_e=-9.864 \times 10^{-7}\,{\rm GeV^9}$, $C_f=(1.214{\rm GeV^2}-0.109T^2_1+0.170T^2_2)\times 10^{-6}\,{\rm GeV^{10}}$, $C_A^{\prime}=(-6.341{\rm GeV^2}+0.064T^2_1-0.075T_2^2)\times 10^{-5}\,{\rm GeV^{10}}$, $C_B^{\prime}=(-4.903{\rm GeV^2}+0.039T_1^2-0.054T_2^2)\times 10^{-5}\,{\rm GeV^8}$, $C_C^{\prime}=(-2.830{\rm GeV^2}+0.023T_1^2-0.031T_2^2)\times 10^{-5}\,{\rm GeV^8}$, and $(C_A\,,C_B\,,C_C)=(C_A^{\prime}\,,C_B^{\prime}\,,C_C^{\prime})\cdot K^T$.
The $3D$ graphs of the $g_{P}^a$ are shown in Figs.\ref{3Dfigg1a}, as for the graphes of other hadronic coupling constants, please consult  the preprint version of the present paper \cite{Present}. One can clearly see  that ``flat surfaces" are obtained for all the hadronic coupling constants, which also presents a reference that it is reasonable to  set $T_1^2=T_2^2$ for simplicity \cite{Decay-mole-WZG-WX,WZG-Z4600-decay,WZG-Pc4312-decay-tetra,DZGWang}.

For the error bounds of the hadronic coupling constants, we follow the approximations   $\frac{\delta \lambda_P}{\lambda_P}=\frac{\delta \lambda_{P^\prime}}{\lambda_{P^\prime}}=\frac{\delta \lambda_\Delta}{\lambda_\Delta}=\frac{\delta \lambda_N}{\lambda_N}=\frac{\delta f_{J/\psi}}{f_{J/\psi}}=\frac{\delta f_{\eta_c}}{f_{\eta_c}}$ to estimate the uncertainties \cite{WZG-Pc4312-decay-tetra,DZGWang}. Need to point out that the uncertainties of the masses of $P_c(4312)$ and $P_c(4330)$ are caused by the input parameters $\langle\overline{q}q\rangle$, $\langle\overline{q}g_s\sigma Gq\rangle$, $m_0^2$, etc \cite{mass-mole-WXW-SCPMA}, so we do not consider their contributions  to  the uncertainties for the strong coupling constants to avoid over evaluation for the uncertainties. What's more, we  abandon the error bounds due to the uncertainties of the free parameters $\delta C_Z$ \cite{Decay-mole-WZG-WX,WZG-Z4600-decay,WZG-Pc4312-decay-tetra,DZGWang}. Since it is not clear to show the error bounds on the $3D$ graphs, we draw the curves of the uncertainties of the diagonals of the Borel platforms in Figs.\ref{3Dfigg1a} for $g_{P}^a$ (the other graphes can be found in the preprint version \cite{Present}) and extract the numerical results in Table \ref{Borel-mass}, it is clear that the strict constraint condition $\chi_Z<6\%$ holds for all the Borel platforms. Accordingly, we obtain the partial decay widths,
\begin{eqnarray} \label{Par-width}
\Gamma^{a} (P_c(4312)\rightarrow \eta_c N) &=& 0.11^{+0.03}_{-0.03} \,{\rm MeV}\,, \nonumber\\
\Gamma^{b} (P_c(4312)\rightarrow \eta_c N) &=& 0.11^{+0.02}_{-0.02} \,{\rm MeV}\,, \nonumber\\
\Gamma (P_c(4312)\rightarrow J/\psi N) &=& 10.78^{+8.97}_{-8.09} \,{\rm MeV}\,, \nonumber\\
\Gamma^{e} (P_c(4330)\rightarrow \eta_c \Delta) &=& 0.10^{+0.01}_{-0.01} \,{\rm MeV}\,, \nonumber\\
\Gamma^{f} (P_c(4330)\rightarrow \eta_c \Delta) &=& 0.10^{+0.06}_{-0.06} \,{\rm MeV}\,, \nonumber\\
\Gamma (P_c(4330)\rightarrow J/\psi \Delta) &=& 57.86^{+33.00}_{-32.90} \,{\rm MeV}\,,
\end{eqnarray}
where the superscripts $a$, $b$, $e$ and $f$ denote the hadronic coupling constants $g_{P}^a$, $g_{P}^b$,  $g_{P'}^e$ and $g_{P^\prime}^f$ have been chosen, respectively. From Eq.\eqref{Par-width}, we can see explicitly that the partial decay widths using the hadronic coupling constants $g_P$ and $g_{P'}$ from different QCD sum rules are consistent with each other, which indicates that the calculations are reliable. The partial decays $P_c(4312)\rightarrow J/\psi N$ and $P_c(4330)\rightarrow J/\psi \Delta$ almost satisfy the total decay widths, and the total decay width $\Gamma_{P_c(4312)}=10.89^{+8.97}_{-8.09} \,{\rm MeV}$ is in very good agreement with the experimental data $9.8\pm2.7^{+3.7}_{-4.5}\,\rm{MeV}$ from the LHCb collaboration \cite{RAaij2}, and supports assigning the $P_c(4312)$ as the $\bar{D}\Sigma_c$ molecular state having the isospin $|I,I_3\rangle=|\frac{1}{2},\frac{1}{2}\rangle$,
the observation of its isospin cousin $P_c(4330)$ with the $|I,I_3\rangle=|\frac{3}{2},\frac{1}{2}\rangle$ in the $J/\psi \Delta$ mass spectrum could shed light on the nature of the $P_c$ states.

As the decay mode $\Delta \to N\pi$ accounts for $99.4\%$ of the total width of the $\Delta$ from the Particle Data Group \cite{PDG}, we expect that the partial decay widths have the relations $\Gamma(P^\prime\to J/\psi \Delta)=\Gamma(P^\prime\to J/\psi N\pi)$ and
$\Gamma(P^\prime\to \eta_c \Delta)=\Gamma(P^\prime\to \eta_c N\pi)$. Then we take account of the finite width effects of the $\Delta$, and obtain the partial decay widths,
\begin{eqnarray} \label{Par-width-Delta}
\Gamma(P^\prime\to \eta_c N\pi)&=&\frac{1
}{16\pi^2 m_{P^\prime}^2  } \int_{(m_N+m_\pi)^2}^{(m_{P^\prime}-m_{\eta_c})^2}ds\,|T_{\eta_c}|^2\frac{\sqrt{s}\,\Gamma_{\Delta}(s)\,
p(m_{P^\prime},m_{\eta_c},\sqrt{s})}{(s-m_{\Delta}^2)^2+s\Gamma_{\Delta}^2(s)}\ , \nonumber\\
&=& 0.057^{+0.008}_{-0.007} \,\rm{MeV}\, ,\nonumber\\
\Gamma(P^\prime\to J/\psi N\pi)&=&\frac{1
}{16\pi^2 m_{P^\prime}^2  } \int_{(m_N+m_\pi)^2}^{(m_{P^\prime}-m_{J/\psi})^2}ds\,|T_{J/\psi}|^2 \frac{\sqrt{s}\,\Gamma_{\Delta}(s)\,p(m_{P^\prime},m_{J/\psi},\sqrt{s})}{(s-m_{\Delta}^2)^2+s\Gamma_{\Delta}^2(s)}\ , \nonumber\\
&=&93.87^{+52.91}_{-51.12} \,\rm{MeV}\, ,
\end{eqnarray}
where
\begin{eqnarray}
|T_{J/\psi}|^2&=&\Sigma \,
|\langle J/\psi(p)\Delta(q)|\mathcal{P}_{P^\prime}(p^\prime)\rangle|^2\, , \nonumber\\
|T_{\eta_c}|^2&=&\Sigma\,
|\langle \eta_c(p)\Delta(q)|\mathcal{P}_{P^\prime}(p^\prime)\rangle|^2\, , \nonumber\\
\Gamma_{\Delta}(s)&=& \Gamma_{\Delta}(m_{\Delta}^2)\frac{p(\sqrt{s},m_N,m_\pi)^3}{p(m_\Delta,m_N,m_\pi)^3}\, ,
\end{eqnarray}
$p^\prime=q+p$, $p(A,B,C)=\frac{\sqrt{\left[A^2-(B+C)^2\right]\left[A^2-(B-C)^2\right]}}{2A}$, $p^{\prime2}=m_{P^\prime}^2$, $p^2=m^2_{J/\psi}$ or  $m^2_{\eta_c}$, $q^2=s$ and
 $\Gamma_{\Delta}(m_{\Delta}^2)=117\,\rm{MeV}$ \cite{PDG}. The partial decay width $\Gamma(P^\prime\to J/\psi N\pi)$ ($\Gamma(P^\prime\to \eta_c N\pi)$) is greatly amplified (diminished) when the finite  width of the $\Delta$ is included. We can estimate the total width by  the dominant decay mode,   $\Gamma_{P_c(4330)}=93.87^{+52.91}_{-51.12} \,\rm{MeV}$, which is compatible with our naive expectation that the resonant states  have widths of the order of hundred  MeV.

The masses of the $\bar{D}\Sigma_c$ molecular states with the isospins  $I=\frac{1}{2}$ and $\frac{3}{2}$ are
$4.31^{+0.07}_{-0.07}\,\rm{MeV}$ and $4.33^{+0.09}_{-0.08}\,\rm{GeV}$, respectively \cite{mass-mole-WXW-SCPMA}, the $P_c(4312)$ can be assigned as the $\bar{D}\Sigma_c$ molecular state with the isospin $I=\frac{1}{2}$, while the molecular state with the $I=\frac{3}{2}$ (or $P_c(4330)$) has not been observed yet. For example, the uncertainty of the mass  $\delta={}^{+0.09}_{-0.08}\,\rm{GeV}$ leads to the uncertainties  ${}^{+196.04}_{-90.37}\,\rm{MeV}$ and  ${}^{+0.232}_{-0.056}\,\rm{MeV}$ for the partial decay widths $\Gamma(P^\prime \to J/\psi \Delta)$ and $\Gamma(P^\prime \to \eta_c \Delta)$, respectively, which are too large, and we discard them, as the central value of the mass of the $P_c(4330)$ corresponds to that of the $P_c(4312)$, and we take the central values in calculating the decay widths. According to calculations of the QCD sum rules, the uncertainties of the masses and pole residues also come from the uncertainties of the input parameters at the QCD side, including both the uncertainties of the hadron masses and hadronic coupling constants suffer from doubling counting.

In Ref.\cite{Decay-mole-WZG-WX}, we choose the current which does not have definite isospin, there exist contributions come from both the $I=\frac{1}{2}$ and $\frac{3}{2}$ molecular states with the $J^P={\frac{1}{2}}^-$, as the mass-gap between the $I=\frac{1}{2}$ and $\frac{3}{2}$ molecular states are vary small, and they have almost degenerated pole residues, so it cannot make much difference by assuming they are the same particle,  if only numerical values are concerned. For the partial decay widths of the $P_c(4312)\to \eta_c N$ and $J/\psi N$, the predictions in Ref.\cite{Decay-mole-WZG-WX} and in the present work are almost degenerated. However, we  should bear in mind that the $P_c(4312)$ and $P_c(4330)$ are two different particles and have different isospins, even if they have degenerated masses and pole residues, we still want to distinguish them and obtain unpolluted predictions.

In Ref.\cite{WZG-Pc4312-decay-tetra},  we take the $P_c(4312)$ as  the  diquark-diquark-antiquark type hidden-charm  pentaquark state with the  $J^P={\frac{1}{2}}^-$, and investigate  the partial decay widths with the QCD sum rules, and observe that the partial decay widths of the $P_c(4312)\to \eta_c N$ and $J/\psi N$ are comparable,  which differ from the predictions in Ref.\cite{Decay-mole-WZG-WX} and in the present work greatly. We can search for the decay mode $P_c(4312)\to \eta_c N$ and precisely measure the branching fractions to distinguish  the pentaquark and molecule assignment.

\section{Conclusions}
In the present study, we investigate the strong decays of the $P_c(4312)$ and its possible higher isospin cousin $P_c(4330)$  considering  conservation of the isospin.  The hadronic coupling constants in the four decay channels $P_c(4312) \rightarrow \eta_c+N$, $ J/\psi+N$, $P_c(4330)\rightarrow \eta_c+\Delta$ and $ J/\psi+\Delta$ are calculated in detail via the QCD sum rules. Then we obtain the partial decay widths, among which, the partial decays $P_c(4312)\rightarrow J/\psi N$ and $P_c(4330)\rightarrow J/\psi \Delta$ almost saturate the total decay widths, and the  width $\Gamma_{P_c(4312)}=10.89^{+8.97}_{-8.09} \,{\rm MeV}$ is in very good agreement with the experimental data $9.8\pm2.7^{+3.7}_{-4.5}\,\rm{MeV}$ from the LHCb collaboration, and supports assigning the $P_c(4312)$ as the $\bar{D}\Sigma_c$ molecular state with the isospin $|I,I_3\rangle=|\frac{1}{2},\frac{1}{2}\rangle$,
the observation of its isospin cousin $P_c(4330)$ with the $|I,I_3\rangle=|\frac{3}{2},\frac{1}{2}\rangle$ in the $J/\psi \Delta$ mass spectrum could shed light on the nature of the $P_c$ states.

\section*{Acknowledgements}
This work is supported by National Natural Science Foundation, Grant Number 12175068 and the Fundamental Research Funds for the Central Universities of China.

\end{document}